\documentclass[12pt]{iopart}

\usepackage{graphicx}
\usepackage[numbers,sort&compress]{natbib}
\usepackage{amsmath}
\usepackage{amsfonts}
\usepackage{amssymb}
\usepackage{graphicx}
\usepackage{color}
\usepackage{txfonts}
\usepackage[titletoc]{appendix}
\usepackage[colorlinks={true}]{hyperref}
\hypersetup{citecolor={blue}, filecolor={blue}, linkcolor={blue}, urlcolor={blue}}

\begin{document}

\title{Pulse-area theorem for precision  control of the rotational motions of a single molecule  in a cavity}

\author{Li-Bao Fan$^{1}$, Chuan-Cun Shu$^{1, *}$}
\address{$^1$ Hunan Key Laboratory of Nanophotonics and Devices, Hunan Key Laboratory of Super-Microstructure and Ultrafast Process, School of Physics, Central South University, Changsha 410083, China}

\ead{\mailto{cc.shu@csu.edu.cn}}
\vspace{10pt}
\begin{indented}
\item[]
\end{indented}

\begin{abstract}
We perform a combined analytical and numerical investigation to explore how an analytically designed  pulse can precisely control the rotational motions of a single-molecular polariton formed by the strong coupling of two low-lying rotational states with a single-mode cavity. To this end, we derive a pulse-area theorem that gives amplitude and phase conditions of the pulses in the frequency domain for driving the polariton from a given initial state to an arbitrary coherent state. The pulse-area theorem is examined for generating the maximum degree of orientation using a pair of pulses. We show that the phase condition can be satisfied by setting the initial phases of the two identically overlapped pulses or by controlling the time delay between pulses for practical applications. 
\end{abstract}

\section{Introduction}
Exploring strong-light-matter coupling has attracted much attention in different fields over the last three decades \cite{PRL2015, JPCL2016NaI, PNAS2017, JPCL2018Na, JPCL2018, JPCL2020, PRB20212}. By exploiting strong coupling between molecules and confined electromagnetic field modes, polaritonic chemistry is emerging as a new interdisciplinary field  \cite{cs2018, LPR2019,csr2019, PRX2020, PNAS-2021,science-Ebbesen}. Considerable theoretical and experimental works have demonstrated that even using a vacuum cavity can modify the energy landscape of molecules and the underlying transitions between states by mixing photonic characters into molecules,  affecting the rates and yields of chemical reactions, emission properties, electronic and excitonic transport, and more  \cite{science2019,pnas2020,science2020,nc2021, JACS2021, JCP2021}. As a result, molecular polaritons as hybrid quasi-particles exhibit many novel chemical and physical phenomena beyond bare molecules      \cite{PRX2015, PRL2019, JCP2020, PRB2021, PRL20212, JCP20212}. This, in turn, leads to molecular polariton as a new platform for studying strong light-matter interactions at the molecular level.   \\ \indent
  Unlike atoms, molecules possess various internal vibrational and rotational degrees of freedom, resulting in the complexity of molecular electronic states with vibrational and rotational fine structures  \cite{Schoelkopf2006, DeMille2008, Pupillo2009, Shuman2010, Rellergert2013}. The rotational energy levels occupy the low-energy part of the energy spectrum. The rotational dynamics of low-lying rotational states can be described well in the framework of rigid-body approximation by considering isolated molecules in the electronic and vibrational ground state  \cite{Barone2019, Shupra2020, Shupra2021}. The interaction Hamiltonian with external fields usually is established within the semiclassical approximation, which treats the field as the classical electromagnetic field and the molecule quantum mechanics  \cite{JCM1963, William1978, Ruggenthaler2018}. Because of its potential applications in chemical physics, quantum information, and quantum simulation,  these advantages of rotational molecules have inspired many theoretical and experimental works to explore quantum control of molecular rotation  \cite{machholm2001,shu2008,qr-TS, IRPC2010, IJC2012, qr-sugny, nc20192,  CEJ2017}. It naturally raises a fundamental question of how to explore the rotational dynamics of molecule-photon hybrid systems driven by externally applied electromagnetic fields and control over molecular polariton from an initial state to a desired target state. \\ \indent
Most recently, we contributed a theoretical proposal to complete coherent control of molecular polariton by considering two-rotational states of a single molecule strongly coupled with a single-mode cavity and driven by pulses \cite{Shu2023}. We showed how to analytically design a composite pulse capable of generating the maximal orientation of the molecular polariton. In this work, we extend the pulse-area theorem to design control pulses that can drive the single-molecular polariton from a given polariton state to an arbitrary coherent state. We show that this goal can be obtained by analytically designing the amplitudes and phases of a composite pulse and how the relative phase and the time delay between the pulses can be used as control parameters in practical applications. This work provides a  theoretical framework for studying the precision control of molecules in a cavity. It has potential applications in quantum optics, polariton chemistry, quantum computing, and simulation. \\ \indent 

We organize the following paper as follows. In Section \ref{Sec:Model}, we explain the theoretical model and derive the corresponding pulse-area theorem. We then perform the numerical simulations to examine the derived pulse-areas theorem in Section \ref{Sec:Result}. Finally, we summarize the results in Sec.  \ref{Sec:conclusion}.
\section{Theoretical methods \label{Sec:Model}}
\begin{figure}[h]\centering
\resizebox{0.7\textwidth}{!}{%
\includegraphics{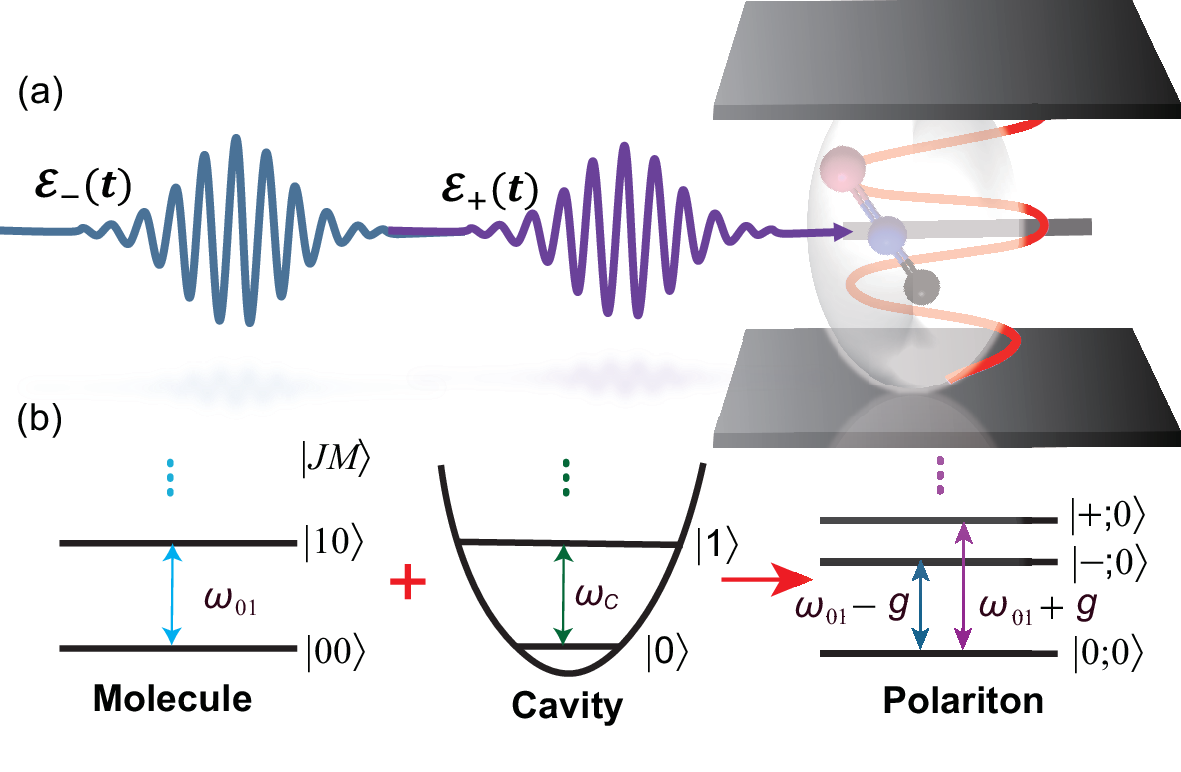} }\caption{(a) Schematic of a composite pulse
-driven molecular-polariton and the corresponding  energy levels.  (b) A single molecule with rotational states $|JM\rangle$ at a rotational frequency $\omega_{01}$  in is strongly coupled with a single-mode cavity with photon states $|n\rangle$ at a frequency $\omega_c$, resulting in entangled states $|0;0\rangle$ and $|\pm;n\rangle$.}
\label{fig1}
\end{figure}
As illustrated in Fig. \ref{fig1}, we consider a single-mode cavity that strongly couples with the two lowest rotational states of a single molecule initially in the ground vibrational level of the ground electronic state, resulting in a rotational molecular polariton. We then apply a time-dependent pulse $\mathcal{E}(t)$ to the polariton system. The corresponding time-dependent Hamiltonian reads $(\hbar=1)$
\begin{eqnarray} \label{H}
\hat{H}(t)=\omega_{c}a^{\dagger}a+B\hat{J}^2-\sqrt{\frac{\omega_c}{2\epsilon_0 V}}\mathbf{\hat{\mu}}\cdot\mathbf{\hat{e}}\left(\hat{a}+\hat{a}^\dag\right)-\mu\mathcal{E}(t)\cos\theta
\end{eqnarray}
where the first term denotes the cavity field Hamiltonian with $a^{\dagger }$ ($a$) being the photon creation (annihilation) operators, $\omega_{c}$ is the frequency of the cavity mode. The second term is the field-free rigid-rotor  Hamiltonian with the angular momentum operator $\hat{J}$ and rotational constant $B$. The third term describes the interaction term between the molecular dipole moment and the cavity,  where $\epsilon_{0}$  the electric constant, $V$ the volume of the electromagnetic mode, and $\hat{e}$ the polarization vector of the cavity mode. The last term describes the interaction between the molecular dipole moment and the laser field by considering the pulse polarization along the same polarization of the cavity, where $\theta$ the angle between the electric field polarization vector and the transition dipole vector.

In the strong-coupling regime, it remains challenging to obtain the time evolution of the system governed by the Hamiltonian in Eq. (\ref{H}). By considering the case that the transition frequencies  $\omega_{01}$ of the molecular system  is in resonance with cavity-photon frequency $\omega_c$ and the coupling strength $g$ is significantly smaller than the frequency $\omega_{01}$, the first three terms of the Hamiltonian in Eq. (\ref{H}) can be written in  Jaynes Cummings model as 
\begin{eqnarray}
\hat{H}_{\mathrm{JC}}=\omega_{01}|10\rangle\langle10|+\omega_c\hat{a}^\dag\hat{a}+g(|00\rangle\langle10|\hat{a}^\dag+\hat{a}|10\rangle\langle00|)
\end{eqnarray}
where  the total excitation number operator $\hat{N}=\hat{a}^{\dagger}\hat{a}+|10\rangle\langle10|$ that is a conserved quantity based on the commutative relation $[\hat{N},\hat{H}_{\mathrm{JC}}]=0$, and $g=\sqrt{\omega_c/(2\epsilon_0 V)}\langle 00|\mu\cos\theta|10\rangle$ with $\mu_{01}=\langle 00|\mu\cos\theta|10\rangle$ the transition dipole moment.  By diagonalizing the  Hamiltonian $\hat{H}_{\mathrm{JC}}$, we obtain the eigenvalues  of the molecular polariton
\begin{eqnarray}
\omega_{0,0}&=&0,\nonumber\\
\omega_{\pm,n}&=&\omega_c(n+1)\pm g\sqrt{n+1}
\end{eqnarray}
and the corresponding  entangled eigenstates
\begin{eqnarray}
|0;0\rangle&=&|00\rangle|0\rangle,\nonumber\\
|\pm;n\rangle&=&\sqrt{2}/2\left(|00\rangle|n+1\rangle\pm|10\rangle|n\rangle\right). 
\end{eqnarray}
We now use the eigenvalues and eigenstates of the polariton to derive the pulsed-driven polariton’s Hamiltonian in hybrid entangled states. The  Hamiltonian of the molecular polariton can be expanded in the representation of entangled eigenstates as 
\begin{eqnarray}
\hat{H}_{\mathrm{JC}}=\sum_{n=0}^{\infty}\sum_{\ell=\pm}\omega_{\ell,n}|\ell;n\rangle\langle \ell;n|.
\end{eqnarray}
The interaction between the molecular dipole moment and the laser field can be written on the entangled polariton basis 
\begin{eqnarray} \label{Hp}
\hat{H}_{\mathrm{p}}(t)&=&-\mathcal{E}(t)\sum_{\ell=\pm}\tilde{\mu}_{0}\Big(|\ell;0\rangle\langle0;0|+|0;0\rangle\langle\ell;0|\Big) \nonumber \\
&&-\mathcal{E}(t)\sum_{n=1}^{\infty}\sum_{\ell,\ell'=\pm}\tilde{\mu}_{\ell}\Big(|\ell;n\rangle\langle\ell';n-1|+|\ell';n-1\rangle\langle\ell;n|\Big),
\end{eqnarray}
where $\tilde{\mu}_0=\pm\sqrt{2}/2\mu_{01}$ and  $\tilde{\mu}_{\pm}=\pm1/2\mu_{01}$ denote the transition dipole moments between entangled states with $\mu_{01}=\langle 00|\mu\cos\theta|10\rangle=\sqrt{3}/3\mu$. The time-dependent wave function $|\psi(t)\rangle$ of the hybrid entangled states can be write as
\begin{eqnarray} \label{WFp}
|\psi(t)\rangle=\sum_{\ell=0,\pm}C_{\ell,0}e^{-i\omega_{\ell,0}t}|\ell;0\rangle+\sum_{n'=1}^{\infty}\sum_{\ell'=\pm}C_{\ell',n'}e^{-i\omega_{\ell',n'}t}|\ell';n'\rangle,
\end{eqnarray}
where $C_{\ell,0}$ denote the complex coefficients of the lowest three states $|0;0\rangle$ and $|\pm;0\rangle$,  and $C_{\pm,n}$  correspond to the complex coefficients of the higher-lying doublet states $|\pm;n\rangle$ with $n>0$. Note that the key technical challenge and novelty of this work were not to extend the JC model to a single two-state molecule but to derive an analytical solution of the pulse-driven quantum JC model, which can be used for describing the rotational dynamics of a single molecule in the cavity.\\ \indent
In this work, we consider the case of $n=0$, and the polariton is initially in the vacuum state $|0; 0\rangle$, which is driven by the fields to an arbitrary coherent state 
\begin{eqnarray}\label{Ts}
|\psi_{\tau_{f}}\rangle=\sum_{\ell=0,\pm}c_{\ell,0}e^{-i(\omega_{\ell,0}t_{f}+\phi_{\ell,0})}|\ell;0\rangle
\end{eqnarray}
where $c_{\ell,0}$ and $\phi_{\ell,0}$ denote the probability amplitude and phase of the state $|\ell; 0\rangle$, respectively.
The fidelity of controlling the quantum system to the target state can be calculated by
\begin{eqnarray}\label{fd}
F =\langle \psi\vert
\psi _{\tau }\rangle\langle \psi _{\tau }\vert\psi\rangle. 
\end{eqnarray}

\subsection{Analytical wave function for a three-state polariton \label{Sec:condition}}
Based on our previous work, the system starting from the initial state $|0;0\rangle$ can be reduced into a three-state system consisting of states $|0;0\rangle$ and $|\pm;0\rangle$ by involving the photon blockade effect \cite{Shu2023}. The corresponding Hamiltonian of the system reads  \cite{Shupra2020, Shupra2021, Guopccp, GuoPRA}
\begin{equation}
H_{I}(t) =\left(
\begin{array}{ccc}
0 & \sqrt{2}/2\mu_{01}\mathcal{E}(t)e^{-i\omega_{-,0}t} & -\sqrt{2}/2\mu_{01}\mathcal{E}(t)e^{-i\omega_{+,0}t}\\
\sqrt{2}/2\mu_{01}\mathcal{E}(t)e^{i\omega_{-.0}t} & 0 &0\\
 -\sqrt{2}/2\mu_{01}\mathcal{E}(t)e^{i\omega_{+,0}t}& 0&0
\end{array}
\right).
\end{equation}
In the interaction picture, the time evolution of the system  from the initial time $t_{0}$ to a given time $t$ can be described by a unitary $\hat{U}(t,t_{0})$
\begin{equation}
\hat{U}(t,t_{0})=\mathcal{I}-i\int_{t_{0}}^{t} d t'H_{I}(t')\hat{U}(t',t_{0}) 
\end{equation}
where $H_{I}(t)=\text{exp}(iH_{0}t)[-\tilde{\mu}_{0}\varepsilon(t)]\text{exp}(-iH_{0}t)$, with the field-free Hamiltonian of the three-level system $H_{0}=\omega_{-,0}|-;0\rangle\langle-;0|+\omega_{+,0}|+;0\rangle\langle +;0|$. 
By involving the Magnus expansion, the time-dependent unitary operator $\hat{U}(t,t_{0})$ can be written as 
\begin{eqnarray}
  \hat{U}(t,t_{0}) =\exp{\left[\sum_{n=1}^{\infty }\hat{S}^{(n)}(t)\right]}  
\end{eqnarray}
where the first three leading terms can be given by means of the Baker-Campbell-Hausdorff formula as $\hat{S}^{(1)}(t)= -i\int_{t_{0}}^{t}dt_{1}H_{I}(t_{1})$, $\hat{S}^{(2)}(t)= (-i)^{2}/2\int_{0}^{t}dt_{1}\int_{0}^{t_{1}}dt_{2}[H_{I}(t_{1}),H_{I}(t_{0})]$, $\hat{S}^{(3)}(t)= (-i)^{3}/6\int_{0}^{t}dt_{1}\int_{0}^{t_{1}}dt_{2}\int_{0}^{t_{2}}dt_{3}\{H_{I}(t_{1}),[H_{I}(t_{2}), H_{I}(t_{2})]\}$. By considering the first-order Magnus expansion, we have
\begin{eqnarray}
\hat{S}^{(1)}(
t)&=& -i\int_{t_{0}}^{t}dt'H_{I}(t')\nonumber\\
&=&i\left(
\begin{array}{ccc}
0 & \theta _{-,0}(t) & \theta _{+,0}(t) \\
\theta _{-,0}^{\ast }(t) & 0 & 0 \\
\theta _{+,0}^{\ast }(t) & 0 & 0
\end{array}
\right),
\end{eqnarray}
where the complex pulse-area  is defined by $\theta_{l,0}(t)=\tilde{\mu}_{0}\int_{t_{0}}^{t}dt^{\prime}\mathcal{E}(t^{\prime})e^{-i\omega_{l}t^{\prime}}$. The first leading term of the unitary operator in the Magnus expansion  can be given by  
\begin{eqnarray}
U^{(1)}(t,t_{0})=\sum_{s=0,\pm}\exp[i\lambda_{s}(t)]
\vert\lambda_{s}\rangle\langle\lambda_{s}\vert
\end{eqnarray}
where $\lambda_{0}(t)=0$, $\lambda_{\pm}(t)=\pm\theta_{0}(t)=\pm\sqrt{\vert\theta_{-,0}(t)\vert^2+\vert\theta_{+,0}(t)\vert^2}$ are the eigenvalues of $\hat{S}^{(1)}(t)$, and the corresponding eigenstates are
\begin{eqnarray}
\vert\lambda_{0}\rangle&=&\left(-\frac{\vert\theta_{+,0}(t)\vert}{\theta_{-,0}(t)}\vert
-;0\rangle +\vert+;0\rangle \right), \nonumber\\
\vert\lambda_{-}\rangle&=&\left(-\frac{\theta _{0}(t)}{\theta_{+,0}^{\ast}(t)}\vert
0;0\rangle+\frac{\theta _{-,0}^{\ast}(t)}{\theta_{+,0}^{\ast}(t)}\vert -;0\rangle+\vert +;0\rangle \right), \nonumber\\
\vert\lambda_{+}\rangle&=&\left(\frac{\theta _{0}(t)}{\theta_{+,0}^{\ast}(t)}\vert
0;0\rangle+\frac{\theta _{-,0}^{\ast}(t)}{\theta_{+,0}^{\ast}(t)}\vert -;0\rangle+\vert +;0\rangle \right).
\end{eqnarray}
The corresponding wave functions in terms of the first-order Magnus expansion can be obtained  by using $\vert\psi_{\vert l;0\rangle}(t)\rangle_{I}=U^{(1)}(
t,t_{0})\vert 0;0\rangle$, i.e., 
\begin{equation}\label{psir}
\vert\psi^{(1)}(t)\rangle=\cos\theta_{0}(t)e^{-i\omega_{0,0}t}\vert 0;0\rangle+i\frac{\theta_{-,0}^{*}(t)}{
\theta_{0}(t)}\sin\theta_{0}(t)e^{-i\omega_{-,0}t}\vert -;0\rangle+i\frac{\theta_{+,0}^{*}(t)}{
\theta_{0}(t)}\sin\theta_{0}(t)e^{-i\omega_{+,0}t}\vert +;0\rangle.
\end{equation}

\subsection{Optimal control conditions for designing control fields}
To entirely transfer the initially state $\vert0;0\rangle$ to the arbitrary target state Eq. (\ref{Ts}), the complex pulse-areas should satisfy the relations
\begin{eqnarray}\label{condition2}
\vert\cos\theta_{0}(t_{f})\vert&=&\vert c_{0,0}\vert\nonumber\\
\left\vert i\frac{\theta_{-,0}^{*}(t_{f})}{\theta_{0}(t_{f})}\sin\theta_{0}(t_{f})\right\vert&=&\vert c_{-,0}\vert\nonumber\\
\left\vert i\frac{\theta_{+,0}^{*}(t_{f})}{\theta_{0}(t_{f})}\sin\theta_{0}(t_{f})\right\vert&=&\vert c_{+,0}\vert
\end{eqnarray}
we can derive that the amplitudes of the complex pulse-areas $\theta_{\pm,0}$ should satisfy the following conditions 
\begin{eqnarray}\label{cd2}
\vert\theta_{\pm;0}\vert=\frac{\vert c_{\pm,0}\vert\arccos{(\vert c_{0,0}\vert)}}{\sqrt{\vert c_{-,0}\vert^{2}+\vert c_{+,0}\vert ^{2}}}.
\end{eqnarray}
 By analyzing the  phases of the states to the target state, we have  
\begin{eqnarray}
\arg[\cos\theta_{0}(t_{f})]&=&\phi_{0}=0,\nonumber\\
\arg\left[i\frac{\theta_{-;0}^{*}(t_{f})}{\theta_{0}(t_{f})}\sin\theta_{0}(t_{f})\right]&=&\phi_{-;0},\nonumber\\
\arg\left[i\frac{\theta_{+;0}^{*}(t_{f})}{\theta_{0}(t_{f})}\sin\theta_{0}(t_{f})\right]&=&\phi_{+;0}.
\end{eqnarray}
It implies that the phases  of  the complex pulse-areas $\theta_{\pm,0}$ in Eq. (\ref{psir}) are required to satisfy the following conditions
\begin{eqnarray}\label{phc}
\arg[\theta_{\pm;0}]=\phi_{\pm;0}-\pi/2.
\end{eqnarray}

\subsection{Designing control fields with optimal amplitudes and phases}
To use the above amplitude and phase conditions in Eqs. (\ref{cd2}) and (\ref{phc}), we make a Fourier transform  of the control field $\mathcal{E}(t)$ to the frequency domain
\begin{eqnarray}\label{eom}
E(\omega)\equiv A(\omega)e^{i\phi(\omega)}=\int_{t_0}^{t_f}dt'\mathcal{E}(t')\exp(-i\omega t'),
\end{eqnarray}
where $A(\omega)$ and $\phi(\omega)$ denote the spectral amplitude and spectral phase of the control field $\mathcal{E}(t)$. From the definitions of the complex pulse-areas $\theta_{\pm}(t_f)$, we can obtain the relations 
\begin{eqnarray}\label{eom1}
\theta_{\pm,0}(t_{f})=\tilde{\mu}_{0}A(\omega_{\pm,0})e^{i\varphi(\omega_{\pm,0})}.
\end{eqnarray}
We can see that the complex pulse-areas of $\theta_{\pm,0} (t_f)$ depend only on the components of the spectral amplitude $A(\omega)$ and phases $\phi(\omega)$ at the transitions frequencies $\omega_{\pm,0}$. Thus,  any shape of control fields can produce the desired target state if their amplitudes and phases meet the conditions. Without loss of generality, we consider two time-delayed control fields in Fig. \ref{fig1} with Gaussian frequency-distributions
\begin{eqnarray} \label{Aw}
E(\omega)=\sum_{\ell=\pm}a_\ell e^{-\frac{(\omega-\omega_{\ell})^2}{2\Delta\omega_\ell^2}}e^{i\varphi_\ell}e^{-i\omega\tau_\ell},
\end{eqnarray}
where $a_\ell$, $\omega_\ell$, $\Delta\omega_\ell$, $\varphi_\ell$ and $\tau_\ell$ denote the amplitude, central frequency, bandwidth, spectral phase, and center time of $\ell$th pulse, respectively.  Furthermore, we consider the two control fields in resonance with the corresponding transition frequencies with $\omega_{\ell}=\omega_{\pm,0}$ and therefore the amplitude conditions in Eq. (\ref{cd2}) can be obtained by setting the values of the amplitudes $a_\pm=\theta_{\pm, 0}/\mu_{\pm}$.  The optimal control fields can be given by 
\begin{eqnarray} \label{At}
\mathcal{E}_o(t)&=&\frac{1}{\pi}\text{Re}\left[{\int_{0}^{\infty}d\omega\sum_{\ell=\pm}\frac{\theta_{\ell,0}}{\mu_{l}} e^{-\frac{(\omega-\omega_{\ell,0})^2}{2\Delta\omega_\ell^2}}e^{i\varphi_\ell}e^{-i\omega\tau_\ell}e^{-i\omega t}}\right],\nonumber\\
&=&\sqrt{\frac{2}{\pi}}\frac{1}{\tau}\left\{\sum_{\ell=\pm}\frac{1}{\mu_{l}}\frac{\vert c_{\ell,0}\vert\arccos(\vert c_{0,0}\vert)}{\sqrt{\vert c_{-,0}\vert^{2}+\vert c_{+,0}\vert^{2}}} e^{-\frac{(t-\tau_{\ell})^2}{2\tau^2}}\cos(\omega_{\ell,0}(t-\tau_{\ell})+\varphi_{\ell})\right\},
\end{eqnarray}
with the pulse duration $\tau=1/\Delta\omega$. By controlling the values of $\varphi_\pm-\omega_{\pm,0}\tau_\pm=\phi_{\pm;0}-\pi/2$, the optimal field in Eqs. (\ref{At}) can satisfy both amplitude and phase conditions by Eqs. (\ref{cd2}) and (\ref{phc}) well as long as the bandwidth narrow enough $\Delta\omega\ll(\omega_{+,0}-\omega_{-,0})$. As a result, we can precisely control the amplitudes and phases of three states $|0;0\rangle$ and $|\pm;0\rangle$ by controlling the values of the amplitudes and phases of the control fields. From quantum optimal control point of view, the pulse-area theorem gives a global optimal solution for the control field, capable of driving the polariton from a given polariton state to an arbitrary coherent state \cite{Sugny2022}.
\section{Application and simulations \label{Sec:Result}}
\subsection{Constructing a coherent superposition as the target state\label{Sec:target}}

To examine this method, we apply it to generate a the maximum degree of orientation for the polariton. By using the target state in Eq. (\ref{Ts}),
the degree of orientation for the molecular polariton  can be written as \begin{eqnarray}\label{cs}
\langle\cos\theta\rangle&=&\langle\psi_{\tau_{f}}(t_{f})\vert\cos\theta\vert\psi_{\tau_{f}}(t_{f})\rangle\nonumber\\
&=&\sum_{\ell=0,\pm}\sum_{s=0,\pm}c_{\ell,0}e^{i((-\omega_{\ell,0}t_{f}+\varphi_{l,0}))}\langle l;0\vert \cos{\theta}\vert s;0\rangle c_{s,0}e^{-i((-\omega_{s,0}t_{f}+\varphi_{s,0}))}\nonumber\\
&=& \vert c_{0,0}\vert\vert c_{-,0}\vert(e^{i((-\omega_{-,0}+\omega_{0,0})t_{f}+(\varphi_{-,0}-\varphi_{0,0}))}+e^{-i((-\omega_{-,0}+\omega_{0,0})t_{f}+(\varphi_{-,0}-\varphi_{0,0}))})\mathcal{M}_{-,0} \nonumber\\
&&+\vert c_{0,0}\vert\vert c_{+,0}\vert(e^{i((-\omega_{+,0}+\omega_{0,0})t_{f}+(\varphi_{+,0}-\varphi_{0,0}))}+e^{-i((-\omega_{+,0}+\omega_{0,0})t_{f}+(\varphi_{+,0}-\varphi_{0,0}))})\mathcal{M}_{+,0} \nonumber\\
&=&2\sum_{\ell=\pm}\vert c_{0,0}\vert\vert c_{\ell,0}\vert\cos(-\omega_{\ell,0}t_{f}+\Delta\phi_{\ell,0})\mathcal{M}_{\ell,0},
\end{eqnarray}
with $\mathcal{M}_{\ell,0}=\langle \ell;0\vert\cos\theta\vert 0;0\rangle$ and $\Delta\phi_{\ell,0}=\phi_{\ell,0}-\phi_{0,0}$.
Based on the method of Lagrange multipliers, the maximum degree of orientation  can be estimated by
\begin{eqnarray}\label{ll}
\mathcal{L}(|c_{0,0}|,|c_{-,0}|,|c_{+,0}|,\lambda)=f-\lambda g,
\end{eqnarray}
with $f=2|c_{0,0}||c_{-,0}|\mathcal{M}_{-,0}+2|c_{0,0}||c_{+,0}|\mathcal{M}_{+,0}$ and $g=\sum_{\ell=0,\pm}|c_{\ell,0}|^{2}-1=0$.
As the maximum of $f$ subject to $g$ is required to  satisfy $\bigtriangledown\mathcal{L}=0$, we have
\begin{eqnarray}\label{mmx}
\begin{aligned}
|c_{-,0}|\mathcal{M}_{-,0}+|c_{+,0}|\mathcal{M}_{+,0}-\lambda |c_{0,0}| &=&0, \nonumber\\
|c_{0}|\mathcal{M}_{-,0}-\lambda |c_{-,0}|&=&0, \nonumber\\
|c_{0}|\mathcal{M}_{+,0}-\lambda |c_{+,0}|&=&0. 
\end{aligned}
\end{eqnarray}
By multiplying each equation in (\ref{mmx}) by $c_{0,0},c_{-,0},c_{+,0}$, we can obtain
\begin{eqnarray}\label{mmxl}
f-\lambda(\vert c_{0,0}\vert^{2}+\vert c_{-,0}\vert^{2}+\vert c_{+,0}\vert^{2})=f-\lambda=0.
\end{eqnarray}
By analyzing  Eqs. (\ref{cs})-(\ref{mmx}),
 the maximum degree of orientation $f$ can be achieved when   $|c_{0,0}|^{2}=|c_{-,0}|^{2}+|c_{+,0}|^{2}$, which corresponds to 
\begin{eqnarray}\label{MDO1}
\lambda &=& \sqrt{\mathcal{M}_{-,0}^{2}+ \mathcal{M}_{+,0}^{2}}=\sqrt{\frac{1}{3}},
\end{eqnarray}
with the conditions $\vert c_{0,0}\vert=\sqrt{2}/2, \vert c_{-,0}\vert=\vert c_{+,0}\vert=1/2$ and $\omega_{-,0}\phi_{+,0}-\omega_{+,0}\phi_{-,0}=2g\pi$. Thus, the target state reads
\begin{eqnarray}\label{psit}
\vert\psi_{\tau}(t_{f})\rangle=\frac{\sqrt{2}}{2}|0;0\rangle +\frac{1}{2}|-;0\rangle e^{i\omega_{-,0}t_{f}+i\phi_{-,0}}+\frac{1}{2}|+;0\rangle e^{i\omega_{+,0}t_{f}+i\phi_{+,0}}.
\end{eqnarray}

\subsection{Numerical simulations}
\begin{figure}[h]\centering
\resizebox{0.8\textwidth}{!}{
\includegraphics{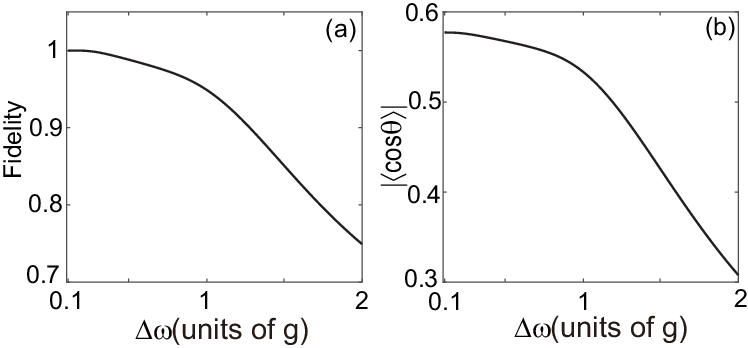}}\caption{Dependence of the numerically calculated (a) fidelity at $t_{f}=50\tau_{0}$  and (b) the degree of orientation at a given time $t=55.278\tau_{0}$ on the bandwidth $\Delta\omega$ with $\tau_{\pm}=0$. The other parameters are chosen with $g=0.1\omega_{01}$, $\varphi_{-}=0$ and $\varphi_{+}=\pi/9$.}
\label{fig2}
\end{figure}

According to Eqs. (\ref{At}) and (\ref{psit} ), the  composite control field can be further given by 
\begin{equation} \label{Atl}
\mathcal{E}_{o}(t)
=\mathcal{E}_{-}e^{-\frac{(t-\tau_{-})^2}{2\tau^2}}\cos\left[\omega_{-,0}(t-\tau_{-})+\varphi_-\right]+\mathcal{E}_{+}e^{-\frac{(t-\tau_{+})^2}{2\tau^2}}\cos\left[\omega_{+,0}(t-\tau_{+})+\varphi_+\right],
\end{equation}
where we can take the strengths of the electric fields with  $\mathcal{E}_{\pm}=\frac{\sqrt{2\pi}}{4\tau\mu_{01}}$ and the phases $\varphi_{\pm}$ with constants to satisfy both amplitude and phase conditions in Eqs. (\ref{cd2}) and (\ref{phc}). Our simulations take  OCS (carbonyl sulfide) molecules at ultralow temperatures as an example with $B=0.20286$ cm$^{-1}$ and $\mu=0.715$ D, which has the rotational period of $\tau_0=\pi/B=82.23$ ps. We consider the molecule initially in the vibrational ground state of the ground electronic state. Since OCS molecules have the fundamental vibrational frequencies of 2174 cm$^{-1}$ for the C-O stretching, 874 cm$^{-1}$ for the C-S stretching, and 539 cm$^{-1}$ for the O-C-S bending, the resonant strong-coupling of the lowest two rotational states in a low terahertz regime and the corresponding driving pulses will not affect the higher rotational and vibrational levels.  We consider the strength of the cavity-coupling $g=0.1\omega_{01}$. \\ \indent
To obtain the time-dependent wave function $|\psi(t)\rangle$ of the system, we numerically solve the time-dependent Schr\"{o}dinger equation governed by the total Hamiltonian $\hat{H}(t)=\hat{H}_{JC}+\hat{H}_{\mathrm{p}}(t)$ without applying the first-order Magnus expansion, in which we use the analytically designed field $\mathcal{E}_o(t)$ in Eq. (\ref{At}) as the control field $\mathcal{E}(t)$ in Eq. (\ref{Hp}). As demonstrated in previous work \cite{Shu2023}, we also examine the model by directly solving Schr\"{o}dinger equation with the time-dependent Hamiltonian in the Rabi model in Eq. (1) without using the first-order Magnus expansion and rotating wave approximation. In addition, we also include higher rotational states of $J=2, 3, 4$ in our simulations. It justifies that the contributions of higher-order terms in the Magnus expansion and higher rotational states can be ignored, and the JC model can be used to describe the strong coupling of $g=0.1 \omega_{01}$ in our results. 

To find an appropriate bandwidth regime of the control field, we first examine the case of keeping the two pulses overlapped with $\tau_{\pm}=0$ and their phases $\varphi_-=0$ and $\varphi_+=\pi/9$. Figure  \ref{fig2} shows the dependence of the fidelity $F$ at $t_{f}=50\tau_{0}$ and the corresponding degree of orientation at $t=55.278\tau_{0}$ on the bandwidth $\Delta\omega=1/\tau$.   We find that the fidelity can reach the value of $F>0.999$ in a narrow bandwidth regime of $\Delta\omega<0.3g$ in Fig. \ref{fig2} (a), for which the maximal orientation corresponds to the value of  $\vert\langle\cos\theta\rangle\vert_{\mathrm{max}}>0.5773$ in Fig. \ref{fig2} (b), in good agreement with the theoretical maximum $\sqrt{1/3}$. Since the two pulses in a broad bandwidth regime will simultaneously open two excitation paths from $\vert 0;0\rangle$ to $\vert \pm;0\rangle$, we can see that the fidelity and the orientation values decrease as the bandwidth increases. In the following simulations, we fix the bandwidth  $\Delta\omega=0.1g$  for further analysis, which can lead to the values of $F=0.9999$ and $\vert\langle\cos\theta\rangle\vert_{\mathrm{max}}=0.57735$. This corresponds to the pulse with the duration of $\tau=1.38$ ns and the intensity at $1.35\times10^9$ W/cm$^2$. This duration of the pulse is much shorter than the rotational decoherence time of the molecules OCS at low temperature.  
\begin{figure}[h]\centering
\resizebox{0.7\textwidth}{!}{
\includegraphics{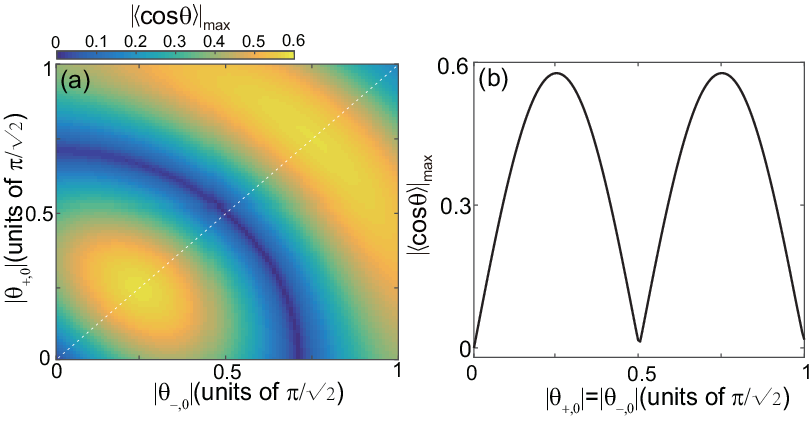}}\caption{Numerical simulations with overlapping control fields having a narrow bandwidth of $\Delta\omega=0.1g$. (a) The maximum degree of orientation $\vert\langle\cos\theta\rangle\vert_{max}$ versus the pulse-areas $\vert\theta_{+,0}\vert$ and $\vert\theta_{-,0}\vert$. (b) $\vert\langle\cos\theta\rangle\vert_{max}$ taken along the line $\vert\theta_{+,0}\vert=\vert\theta_{-,0}\vert$ in (a). The other parameters are the same as those in Fig. \ref{fig2}.}
\label{fig3}
\end{figure}

To show how the amplitude condition by Eq. (\ref{cd2}) determines the degree of orientation, we now examine how the strengths $\mathcal{E}_{\pm}$ can be used to control the populations of three states. Figure \ref{fig3} shows the dependence of the orientation value on the pulse-areas $\theta_{\pm,0}(t_f)$ at $t_f=50\tau_{0}$ by varying the values of $\mathcal{E}_{\pm}$ while keeping other parameters of the pulse the same as those in Fig. \ref{fig2} at $\Delta\omega=0.1g$. 
We can see that the orientation maximum depends strongly on the pulse-areas (i.e., the amplitudes $\mathcal{E}_{\pm}$). As can be seen from Fig. \ref{fig3} (b), the  maximum orientation appears at $\vert\theta_{+}\vert=\vert\theta_{-}\vert=\sqrt{2}\pi/8$ and $3\sqrt{2}\pi/8$, satisfying the amplitude condition in Eq. (\ref{cd2}). It implies that the orientation value at a given time can be controlled by controlling the amplitude values of $\mathcal{E}_{\pm}$.
\begin{figure}[h]\centering
\resizebox{0.7\textwidth}{!}{
\includegraphics{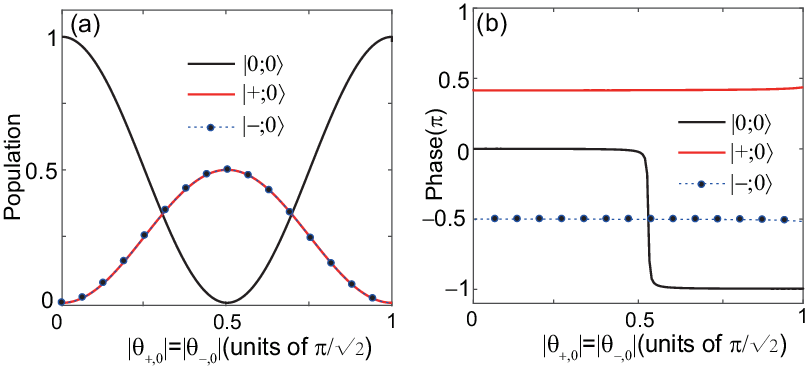}}\caption{Final (a) population and (b) phase of the three states $\vert0;0\rangle$ and $\vert\pm;0\rangle$ as functions of the pulse-areas $\vert\theta_{+,0}\vert=\vert\theta_{-,0}\vert$ at $t_f=50\tau_{0}$. The other parameters are the same as those given in Fig. \ref{fig2}.}
\label{fig4}
\end{figure}
Figures \ref{fig4} (a) and (b) show the populations and phases of the three states versus the amplitude at $t_f=50\tau_{0}$. We can find that the populations of the three states strongly depend on the values of the pulse-areas $|\theta_{\pm,0}|$  and the populations have the distributions of $0.5$ $0.25$ and $0.25$ in three states $\vert 0;0 \rangle $  and $\vert \pm;0 \rangle $ at $\vert\theta_{-,0}\vert=\vert\theta_{+,0}\vert=\sqrt{2}\pi/8$ and $3\sqrt{2}\pi/8$, where the orientation reaches its the theoretical maximum in Fig. \ref{fig3}.   Interestingly, the phases of the excited states $\vert \pm;0 \rangle $ keeps unchanged by varying the values of the pulse-areas, whereas the phase of the ground state $\vert 0;0 \rangle $ flips from 0 to $\pi$ at $\vert\theta_{-,0}\vert=\vert\theta_{+,0}\vert=\sqrt{2}\pi/4$ and keeps unchanged. By analyzing Eq. (\ref{psir}), we can see this phenomenon can be attributed to the sign change of $\cos{\theta}_0(t_f)$. It implies that we can use the the parameters $\mathcal{E}_{\pm}$ of the two pulses to control the phase of the ground state.

\begin{figure}[h]\centering
\resizebox{0.7\textwidth}{!}{
\includegraphics{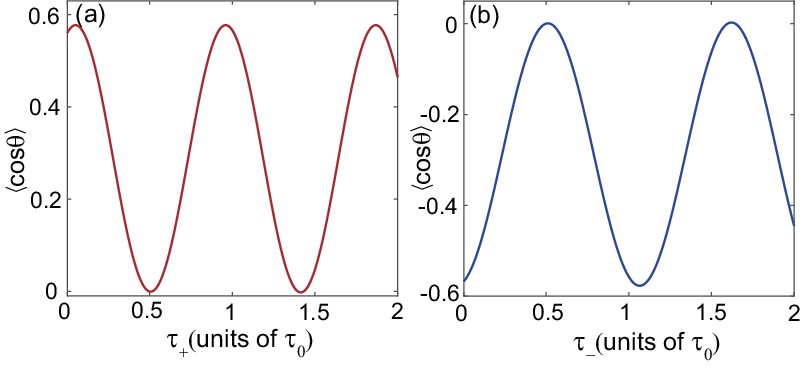}}\caption{Dependence of the numerically calculated degree of orientation on the time delay (a) $\tau_{+}$ at $t=56.389\tau_{0}$ and (b) $\tau_{-}$ at $t=54.773\tau_{0}$ with phase $\varphi_{\pm}=0$. The target state is chosen at $t_f=52.778\tau_{0}$ in (a) and at $t_f=47.727\tau_{0}$ in (b). The other parameter same as in Fig. \ref{fig2}.}
\label{fig5}
\end{figure}

For practical applications, the phases of the pulses can be manipulated by varying the time delay between pulses. To this end, we examine how the scheme depends on the time delay by fixing the phases of the composite pulses at $\varphi_{\pm}=0$. For the first case, we keep the center time  $\tau_{-}=0$ and the other parameters the same as in Fig. \ref{fig2}.   Figure \ref{fig5} (a) shows how the time delay  $\tau_{+}$ affects the observed degree of orientation at $t=56.389\tau_{0}$  in the field-free regime, for which the pulses are turned off at $t_{f}=52.778\tau_{0}$. By analyzing Eq. (\ref{cs}),  we can derive the degree of orientation $\langle\cos\theta\rangle=(1+\cos(\omega_{+,0}\tau_{+}-\pi/9))M_{10}/2$ with the chosen parameters, showing the minimum value of 0 at $\tau_{+}=50\tau_{0}/99+90k\tau_{0}/99$ and the maximum value of  $\sqrt{1/3}$ at $\tau_{+}=5\tau_{0}/99+90k\tau_{0}/99$ with $k=0, 1, 2, \cdots$. We can see that the degree of orientation in Fig.\ref{fig5} (a) varies periodically with the delay time $\tau_+$, in good agreement with the theoretical analysis. We also examine the case in Fig. \ref{fig5} (b) by varying the center time $\tau_{-}$ while keeping $\tau_{+}=0$ and the other parameters the same as those in Fig. \ref{fig5} (a). The change behavior of the orientation in Fig. \ref{fig5} (b) is consistent with the theoretical relation $\langle\cos\theta\rangle=(-1+\cos(\omega_{-,0}\tau_{-}+\pi/11))M_{10}/2$ for the chosen parameters.
\begin{figure}[h]\centering
\resizebox{0.7\textwidth}{!}{
\includegraphics{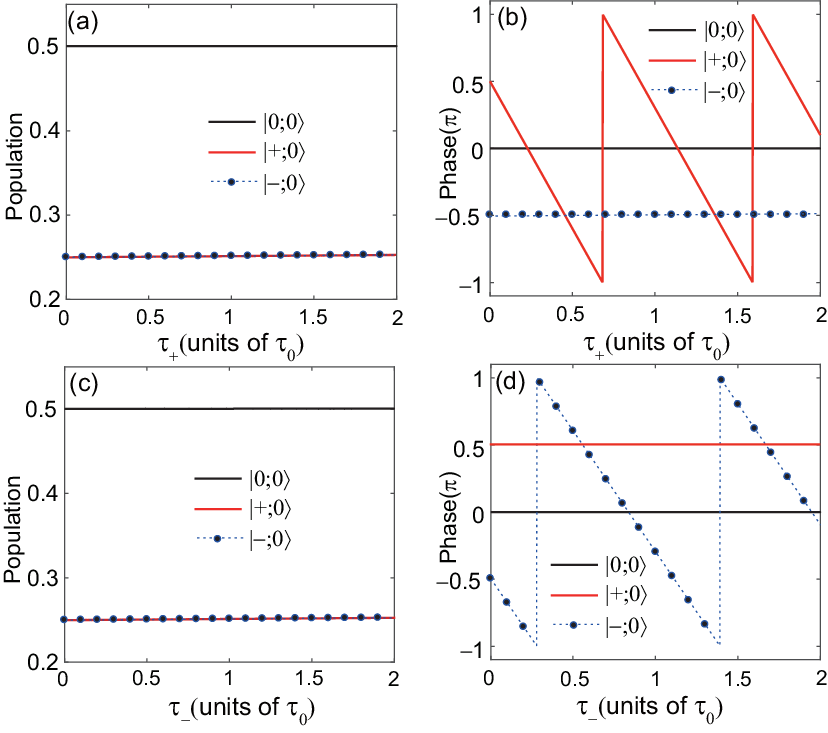}}\caption{The final population and  phase of the three polariton states $\vert0;0\rangle$, and $\vert\pm;0\rangle$ versus the time delay (a) and (b) $\tau_{+}$ at $t_{f}=52.778\tau_{0}$ and (c and d) $\tau_{-}$ at $t_{f}=47.727\tau_{0}$. The other parameter same as in Fig. \ref{fig5}.}
\label{fig6}
\end{figure}

Figure \ref{fig6} shows the corresponding population and phase of the three polariton states versus the time delay  $\tau_{+}$ and $\tau_-$. We can see that the populations in the three states, $\vert 0;0 \rangle $ and $\vert \pm;0 \rangle $, remain unchanged at $0.5$, $0.25$, and $0.25$. The time-delay changes result in the phase changes of the three states, which lead to constructive and  destructive interference phenomena between orientations induced by the pulses $\mathcal{E}_+(t)$ and $\mathcal{E}_-(t)$. Thus, the phases of the three states play crucial roles in leading to the maximum degree of orientation.\\ \indent
Based on these analysis, we can find that the maximum degree of orientation that occurs in the field-free regime depends on the time delay between pulses $\mathcal{E}_+(t)$ and $\mathcal{E}_-(t)$. Using $\tau_+$ or $\tau_-$ as the control parameter will lead to different degree of orientation at a given time in the field-free regime. It implies that the observation of the maximum orientation depends on the time delay of the probe light to the designed control fields in coherent control experiments. The applied pulses that satisfy the control conditions enable whether the designed control schemes can achieve the maximum degree of orientation in principle. 

\section{Conclusion and Outlook \label{Sec:conclusion}}
We have presented a pulse-area theorem for controlling the rotational motions of a single molecule in a cavity, a polariton system formed by the strong coupling of two low-lying rotational states with a single-mode cavity. We considered a pair of pulses that generate a coherent superposition of the lowest three polariton states. We showed how the pulses could be analytically designed by controlling their amplitudes and relative phases to drive the polariton from the ground state to an arbitrarily desired superposition of three states. We performed numerical simulations to examine this method for a single OCS molecule in a cavity. It has been demonstrated that the corresponding polariton could be precisely controlled for generating the maximum degree of orientation. For practical applications, we demonstrated that the derived phase condition of the pulses could be satisfied by controlling the time delay between the pulses. This pulse-area theorem  has  a broad impact and interest across fields in quantum optics, chemical physics, and quantum control, as it offers a solid theoretical proof for controlling molecular rotation in a cavity and has direct applications across fields, e.g., polariton chemistry and molecular polaritonics, for exploring novel quantum optical phenomena \cite{Ultrastrong2019,Ultrastrong2019a}.\\ \indent
This pulse-area theorem as the theoretical proof for exploring the post-pulse orientation at the single molecular-polariton level could inspire experimentalists to develop new strategies toward implementing quantum coherent control of polariton's rotation. The model could be achieved by applying direct laser cooling and magneto-optical trapping to produce an ultracold polar molecule in a single quantum state in a vacuum cavity \cite{barry2014,hemmerling2016,truppe2017,vilas2022}. The experimental implementation of strong coupling in molecules may use cavities, e.g., engineered Fabry-Perot systems or plasmonic structures that require advanced practical fabrication and measurement techniques. Since the coupling strength is inversely proportional to the volume of the cavity, it implies that strong molecule–cavity coupling, even in the low terahertz regime, is accessible by carefully designing the cavity structure and controlling the environment temperature. According to the Boltzmann relationship between temperature and frequency, we can use the inverse of Planck's formula to determine the temperature value of $<$1 K to obtain the initial vacuum photons used in our model \cite{auletta2009,milonni2013quantum}. In addition, the main results, i.e., the amplitude and phase conditions in Eqs. (\ref{cd2}) and (\ref{phc}) can also apply to an ensemble of molecules in a cavity. The observation of a strong coupling effect on the molecular orientation can be examined firstly in the ensemble system by loading multiple molecules into the cavity, which can significantly reduce the coupling strength of the cavity per molecule and, therefore, decrease the challenge in experiments. To this end, the strong and ultrastrong
coupling of collective molecules with the cavity in the terahertz regime has been observed recently in different experiments \cite{damari2019,mavrona2021}.

Our analysis based on the molecule initially in the ground rotational state  $\vert00\rangle$ can be extended to molecules initially in other rotational states. For future studies, it would be interesting to explore the molecule consisting of higher rotational states to increase the maximal degree of orientation while considering the robustness of the proposed control
protocol against experimental errors and limitations, which remains a challenging control problem in the field of quantum control. To address this challenge, optimal and robust control methods, combined with artificial intelligence algorithms, could be employed to search for optimal control fields subject to multiple constraints, leading to a higher degree of orientation of molecules at finite temperature \cite{Sugny2022,sugny2005,Sugny2015,shu2016,shuQuantum2016,guo2018a,guo2019,dong2021a}.
\section*{Acknowledgments}
  This work was supported by the National Natural Science Foundations of China (NSFC) under Grant Nos. 12274470 and 61973317 and the Natural Science Foundation of Hunan Province for Distinguished Young Scholars under Grant No. 2022JJ10070. L.-B. F. acknowledges the financial support in part from the Fundamental Research Funds for the Central Universities of Central South University under Grant No. 1053320211611.
\bibliographystyle{}

\end{document}